# A hybrid SLAM-Payne framework for atmospheric parameter and abundance determination of early-type Stars from LAMOST DR9 low-resolution Spectra

Weijia Sun  [1]⋆

Leibniz-Institut für Astrophysik Potsdam (AIP), An der Sternwarte 16, 14482 Potsdam, Germany

## ABSTRACT

Context. Early-type stars are key drivers of Galactic chemical evolution, enriching the interstellar medium with alpha elements through powerful stellar winds and core-collapse supernovae, fueled by their short lifetimes and high masses. However, their spectra remain challenging to analyse due to rapid rotation, weak metal lines, and non-LTE effects. While large spectroscopic surveys provide extensive low-resolution data, extracting reliable parameters remains difficult due to methodological limitations for hot stars.
Aims. Our goal is to develop a unified framework combining data-driven and synthetic spectral approaches to determine atmospheric parameters and abundances for hot stars using low-resolution spectra, addressing limitations in current methodologies while retaining critical spectral information.
Methods. We present a hybrid approach integrating the Stellar LAbel Machine (SLAM) and the Payne frameworks, for low-resolution ($R \sim 1800$) spectra from LAMOST DR9. Our method preserves full spectral information including Balmer series and metal-line blends, employing neural-network interpolation for efficient parameter estimation ($T_{\rm eff}$, $\log g$, and $v \sin i$) and abundance determination for O, Mg, Si, and Fe, across 8000 K–20,000 K.
Results. We derive stellar parameters and abundances for 315,822 stars with SNR ⩾ 10 in the $r$-band. Among them, we identify 3,564 blue horizontal branch candidates, over 90% of which align with stellar evolutionary models of horizontal branch stars. Additionally, we detect abundance trends ([$\alpha$/Fe] –[Fe/H]) that exhibit temperature-dependent systematics and a distinct $\alpha$-poor stellar population within $0.0 \leqslant$ [Fe/H] $\leqslant 0.5$ dex. The radial abundance gradients are negative and consistent with that derived from Cepheids, with a slope of $-0.070 \pm 0.007$ in [Fe/H] in the region $6 \leqslant R_{\rm GC} \leqslant 15$ kpc.

## 1. Introduction

Early-type stars, with effective temperatures exceeding 7500 K, serve as critical astrophysical laboratories for studying stellar evolution and Galactic chemical enrichment. These stars not only dominate the ultraviolet radiation budget of galaxies (Maeder & Conti 1994), drive chemical mixing through intense stellar winds (Puls et al. 2008), and ultimately explode as supernovae that disperse heavy elements into the interstellar medium (Zinnecker & Yorke 2007), but also are tracers of the young stellar populations within the Galaxy (Venn 1995; Carraro et al. 2010). In particular, understanding the detailed atmospheric parameters and chemical compositions of hot stars is essential for unravelling the complex interplay between stellar evolution and Galactic chemical evolution (e.g., Eldridge & Stanway 2022). The radial metallicity gradient inferred from this young population traces the current state of the disk abundance profile (Bragança et al. 2019). Together with the metallicity profiles of older mono-age populations (Lian et al. 2023), these gradients are important observational constraints for models of Galactic chemical evolution (Maiolino & Mannucci 2019).

Despite their importance, large-scale spectroscopic analyses of hot stars remain challenging due to their rapidly rotating atmospheres, weak metal-line features, and complex non-LTE (NLTE) effects. The spectral analysis of hot stars presents unique challenges compared to their cooler counterparts. High rotation rates broaden absorption features (Royer et al. 2007), while elevated temperatures reduce line densities and enhance NLTE effects (Lanz & Hubeny 2007). These factors, as well as internal mixing processes and the possible presence of chemical peculiarities, complicate the determination of fundamental parameters ($T_{\rm eff}$, $\log g$) and chemical abundances from low-resolution spectra. Traditional methods developed for FGK stars often fail to account for the blended Balmer-line profiles and pressure-broadening effects that dominate hot-star spectra (Takahashi & Langer 2021). Consequently, previous studies of hot stars have largely relied on high-resolution spectroscopy (e.g., Dunstall et al. 2015), limiting sample sizes to $\sim 10^3$ stars despite the availability of $\sim 10^6$ spectra in public surveys like Sloan Digital Sky Survey (SDSS; Almeida et al. 2023) and Large Sky Area Multi-Object Fiber Spectroscopic Telescope (LAMOST; Deng et al. 2012) Although the upgraded spectral resolution of LAMOST's Medium-Resolution Survey (MRS; $R \sim 7500$) offers new opportunities for studying the abundances of hot stars, its limited wavelength coverage constrains the precise determination of surface gravity ($\log g$) and chemical abundances.

Recent advancements in data-driven spectral analysis have opened new avenues for tackling these challenges. Methods such as the Payne (Ting et al. 2019) and Stellar LAbel Machine (SLAM; Zhang et al. 2020a) have demonstrated that it is possible to extract detailed stellar labels even from spectra with moderate resolution. Xiang et al. (2022) introduced the HotPayne method, a spectral analysis tool built upon the Payne but specifically designed for hot stars. Aside from the difference in temperature range, the HotPayne masks the Hydrogen lines to resolve the reverse of spectral gradients for the A0 stars. While this approach successfully mitigates the impact of Hydrogen lines for the A0 stars, an essential part of the information is lost. On the other hand, the SLAM method, building upon Guo et al. (2021); Sun et al. (2021a), has proven effective in deriving parameters. Notably, Sun et al. (2021b); Sun & Chiappini (2024) leveraged LAMOST's MRS DR 9 to catalog over 100,000 late-B- and A-type MS stars, with $T_{eff}$ spanning from 7000 K and 15000 K. They uncovered a bimodal rotation distribution that varies with stellar mass and is influenced by metallicity, suggesting a significant role of chemical composition in the rotational evolution of A-type stars on the main sequence. Recently, the SLAM method has been expanded to include photometric information to constrain the stellar temperature better and has been applied to a subset of Blue horizontal branch (BHB) stars (Ju et al. 2025), selected by the equivalent widths of multiple absorption line profiles (Ju et al. 2024).

Each of these methods has its strengths and limitations; for instance, while SLAM excels in robustness and consistency for temperature determinations, its scalability is limited by slower inference times when handling extensive training sets and high-dimensional label spaces. The Payne framework, on the other hand, enables efficient and flexible abundance determinations across a wide parameter space, but its reliance on the accurate modelling of spectral gradients can lead to reduced reliability in regions where the spectral response is highly non-monotonic. These shortcomings motivate the development of a unified framework that combines the strengths of data-driven and synthetic spectral approaches. This study addresses these challenges through a synergistic analysis of LAMOST Low-Resolution Survey (LRS; $R \sim 1800$) DR9 spectra using complementary methodologies. Building upon our previous work with SLAM (Sun et al. 2021a; Sun & Chiappini 2024), we incorporate Payne's flexible spectral model to enable abundance determinations for four key elements (O, Mg, Si, Fe) across the 8000 K–19,000 K temperature range. Our approach retains the full spectral information content—including the diagnostically critical Balmer series and metal-line blends—while maintaining computational efficiency through neural network interpolation. We validate the method against high-resolution benchmarks and quantify systematics arising from differences in spectral modelling techniques. The resulting catalog of 315,822 stars with $r$-band SNR $\geqslant 10$ represents a homogeneous dataset of hot-star parameters and abundances, providing simultaneous determinations of $T_{eff}$, $\log g$, $v \sin i$, and elemental abundances without excluding broadened features.

The paper is organized as follows: Section 2 describes our source catalogue and pre-selection. Our hybrid SLAM-Payne framework is detailed in Section 3 with a performance validation for precision. Then we present the accuracy test of our measurement in Section 4, via verification against high-resolution data and comparison against low-resolution data. Section 5 presents the stellar parameter catalogue with abundance measurements and identification of distinct stellar populations, including BHB stars and metal-rich outliers. We also discuss the photometric selection and abundance distribution of the sample as well as the abundance gradients across the Milky Way disk. We conclude in Section 6 with a summary of our key findings and implications.

## 2. Data

Our sample of early-type stars is compiled from the low-resolution ($R \sim 1800$) spectra in the ninth data release (DR9) of the LAMOST Experiment for Galactic Understanding and Exploration (LEGUE) surveys (Deng et al. 2012; Zhao et al. 2012; Liu et al. 2014). The LAMOST, a 4 m reflective Schmidt telescope, provides low-resolution spectra with a wavelength coverage of 370 – 910 nm. Its DR9 Version 2.0 spans observations from October 24, 2011 to June 12, 2021, containing 10,809,336 spectra with 10,495,781 classified as stellar spectra. In addition to source catalogs, the release provides a stellar parameter catalog for 6,921,466 spectra of A, F, G and K stars, deriving $T_{eff}$, $\log g$, [Fe/H] and [$\alpha$/M] from the LAMOST Stellar Parameter Pipeline (LASP, Luo et al. 2015).

The primary catalogue of early-type stars is constructed using the source catalogue, supplemented with *Gaia* DR3 and Two Micron All Sky Survey (2MASS; Skrutskie et al. 2006) for photometric determination, and follows Riello et al. (2021)'s guidelines for *Gaia* astrometric and photometric parameters. We adopt the photo-geometric distance estimates from Bailer-Jones et al. (2021), who derived distances using EDR3 parallaxes (corrected for the zero-point offsets in parallax) within a three-dimensional Galactic model.

We apply two selection criteria to refine the early-type candidates:

1. Objects with LASP-derived effective temperatures below 7500 K are excluded, consistent with Sun et al. (2021a). The remaining sample includes stars above this threshold and those lacking LASP temperature estimates, e.g., early-type stars and M/K-type dwarfs (see Figure 2 in Sun et al. 2021a).
2. Using the approach proposed in Zari et al. (2021), we utilise the proxy of the absolute magnitude of a star in the $K_s$ band, denoted as $\tilde{M}_{K_s}$. We set a limit for the cool end of the spectral types, requiring that $\tilde{M}_{K_s}$, defined as $K_s - 5\log_{10}(r) + 5$, be smaller than 2.5 mag. No additional selections based on colours (e.g., $J - K_s$ or $G - K_s$) are applied. This magnitude cut roughly corresponds to the absolute magnitude of an F6V star with $T_{eff} = 6350$ K (Pecaut & Mamajek 2013).

Either of the criteria removes approximately 60% of the candidates from the source catalogue. When combined, these criteria leave 1,833,153 spectra of 1,405,917 stars.

## 3. Method

### 3.1. Pipeline

One of the major difficulties in determining spectral labels from early-type stars' spectra originates from the non-monotonic trend of hydrogen line strength. The equivalent widths of the Balmer lines reach a maximum around 9000 K due to the relatively high excitation energy of the Balmer series (10.2 eV) (Gray & Corbally 2009). In late-type stars, hydrogen lines are weak but steadily stronger with increasing effective temperature until reaching a maximum in early A-type stars. At higher temperatures (B- and O-type stars), the hydrogen lines again become weaker. As the dominant feature for effective temperature diagnostics, this near-zero gradient in hydrogen line strength creates an inflection point around $T_{eff} \sim 9000$ K, making temperature estimates sensitive to initial values and potentially creating artifacts of temperature void around the peak.

This effect has been observed in several spectral analysis methods, including the modules described in Blomme et al.

(2022), where different analysis nodes for hot-star spectra in the Gaia-ESO survey were investigated. While $\chi^2$ minimisation methods are generally robust, machine-learning approaches may be affected by this issue due to the sign change in spectral gradients around A0-type stars. To address this, Xiang et al. (2022) devised a spectral masking technique to exclude hydrogen lines during label determination for early A-type and late B-type stars. However, masking hydrogen lines reduces spectral information content, leading to lower precision in determining $T_{eff}$ and $\log g$. They combined results from analyses with and without masked hydrogen lines to produce final estimates, which resolved the spurious absence of A0-type stars and unexpected Kiel diagram patterns around 10000 K (Xiang et al. 2022), though at the cost of information loss for stars near A0-type.

The SLAM method solves this problem through several advantages inherent to its data-driven approach using Support Vector Regression (SVR). Unlike gradient-based methods, SVR employs dual optimisation with Lagrange multipliers through quadratic programming, avoiding direct gradient calculations of loss functions. The Karush-Kuhn-Tucker conditions in SVR's dual form provide stationary point solutions analogous to gradient-based optima but without gradient descent. Additionally, SLAM automatically determines hyperparameters (penalty level and tube radius) through the training set itself, enabling pixel-wise adaptive model complexity that particularly benefits hydrogen line analysis around A0 stars. This capability was demonstrated in our previous work (Sun et al. 2021a), where SLAM analysis of LAMOST MRS showed no anomalous temperature gaps (see their Fig. 9). While $\log g$ accuracy from MRS remained limited due to the restricted wavelength coverage (5000–5300 Å and 6350–6800 Å), we anticipate improvement with the broader coverage of LRS.

We complement SLAM with an Artificial Neural Network model based on the Payne algorithm to estimate additional abundances. This hybrid approach addresses SLAM's computational limitations as Support Vector Machine complexity scales between $O(n_{features} \times n_{samples}^2)$ and $O(n_{features} \times n_{samples}^3)$ (Chang & Lin 2021), making comprehensive label coverage impractical. Our implementation uses SLAM for four fundamental parameters ($T_{eff}$, $\log g$, [Fe/H], and $v \sin i$) while employing the Payne model for other abundances using an extensive training library with complete label space coverage.

Our pipeline proceeds as follows: First, we obtain best-fit labels from SLAM. These parameters then fix $T_{eff}$, $\log g$, [Fe/H], and $v \sin i$ in a subsequent Payne module run to determine remaining abundances. We report associated uncertainties using standard propagation methods.

For spectral preprocessing, we normalise spectra and correct for radial velocity using `laspec`[1] (Zhang et al. 2021). The normalisation iteratively fits a spline function three times, rejecting pixels deviating $> 3\sigma$ from median values in 100 Å windows. We adopt line-of-sight velocity corrections from LAMOST pipeline-reduced spectral headers.

Section 3.2 describes our template grid and synthetic spectra. Training procedures for SLAM and Payne modules appear in Sections 3.3 and 3.4, respectively, followed by performance evaluation in Section 3.5.

## 3.2. Grid synthesis

We construct a grid comprising $\mathcal{N}$ sets of stellar labels in 11D parameter space ($T_{eff}$, $\log g$, [Fe/H], microturbulence $v_{mic}$, $v \sin i$,

| Label ($l$) | Range | Unit | Sampling |
|---|---|---|---|
| $T_{eff}$ | (7000, 20000) | K | EEP [a] |
| $\log g$ | (0, 5) | dex | EEP |
| [Fe/H] | (−2.5, 1.0) | dex | EEP |
| $v_{mic}$ | (0, 10) | km s$^{-1}$ | Uniform |
| $v \sin i$ | (5, 450) | km s$^{-1}$ | Log-normal [b] |
| [C/Fe] | (−2.5, 1.0) | dex | Uniform |
| [O/Fe] | (−1.5, 1.5) | dex | Uniform |
| [S/Fe] | (−1.5, 1.5) | dex | Uniform |
| [Mg/Fe] | (−1.0, 1.0) | dex | Uniform |
| [Ca/Fe] | (−1.5, 1.0) | dex | Uniform |
| [Si/Fe] | (−1.5, 1.0) | dex | Uniform |

[a] Equivalent evolutionary phase (EEP) in MIST isochrones (Paxton et al. 2011; Choi et al. 2016)
[b] Log-normal distribution from Sun et al. (2021a)

Table 1: Range and sampling method of the stellar labels for training.

[C/Fe] , [O/Fe] , [S/Fe] , [Mg/Fe] , [Ca/Fe] , and [Si/Fe] ). The label selection follows recommendations from Xiang et al. (2022), who used Cramér-Rao bounds to assess theoretical precision limits for hot-star label determination from LAMOST LRS. For $T_{eff}$ and $\log g$, we sample MIST isochrones (Paxton et al. 2011; Choi et al. 2016) using equivalent evolutionary phase (EEP) via `isochrones` (Morton 2015). To ensure smooth label distributions, we perturb these values with normal distributions ($\sigma = 250$ K for $T_{eff}$, $\sigma = 0.25$ dex for $\log g$), then restrict to $T_{eff} \in [7000, 20000]$ K and $\log g \in [0.0, 5.0]$ dex.

We sample [Fe/H] uniformly between −2.5 and 1.0 dex, while other abundances follow uniform distributions scaled relative to [Fe/H] using the solar abundance scale of Asplund et al. (2009). Microturbulence $v_{mic}$ ranges uniformly from 0 to 10 km s$^{-1}$. For $v \sin i$, we adopt the log-normal distribution from Sun et al. (2021a) with peak at 90 km s$^{-1}$ and tail extending to ~ 400 km s$^{-1}$. Table 1 summarises the training grid.

We generate synthetic spectra using SYNTHE with Kurucz LTE atmospheric models from ATLAS12 (Kurucz 2005). Broadening by microturbulence and rotation is applied via `vidmapy`[2] at $R = 100,000$. To match LAMOST-LRS instrumental characteristics, we apply wavelength-dependent line spread functions (LSF) determined by comparing high-SNR ($> 100$) FGK star observations to unbroadened synthetics. The wavelength-varying resolution is approximated by fifth-order polynomials across the blue (3700–5800 Å) and red (5700–9100 Å) arms. We add Gaussian noise equivalent to SNR = 1000 per pixel.

While ATLAS12 assumes LTE, Przybilla et al. (2011) demonstrated LTE models remain adequate for main-sequence stars below 22000 K, justifying our 20000 K upper limit. Synthetic spectra are computed in absolute flux units, including continuum and line contributions. After applying instrumental effects, we resample these spectra to LRS coverage (1 Å per pixel) and normalize them using the iterative spline-fitting procedure described in Section 3.1, which operates on the total flux and reject $> 3\sigma$ outliers in 100 Å windows. Observed raw spectra undergo the same resampling and normalisation process to ensure consistency.

We adopt the line list from the Superfast Line Profile Variability (LPV) project[3], which combines OBA star line lists

---

[1] https://github.com/hypergravity/laspec

[2] https://github.com/RozanskiT/vidmapy
[3] http://www.astro.spbu.ru/LPV/?q=line-list

from Reader et al. (1980), Striganov, A.-R. & Sventitskii, N.-S. (2013), and the NIST database[4].

### 3.3. SLAM module

The SLAM module follows the methodology outlined in Sun et al. (2021a), employing two specialized models: (1) a high-resolution converter mapping synthetic spectra to fundamental parameters ($T_{eff}$, $\log g$, [Fe/H]) within predefined ranges, and (2) a mock spectrum generator producing LRS-like spectra through rotational and instrumental broadening. We mask known diffuse interstellar bands (Hobbs et al. 2008) and exclude the H$\alpha$ region (6552.0–6572.0 Å) to avoid emission line contamination.

Hyperparameter optimisation uses $\epsilon = 0.05$ (tube radius), $C \in \{0.1, 1, 10\}$ (penalty), and $\gamma \in \{0.1, 1, 10\}$ (RBF kernel width), selected via five-fold cross-validation. Computational constraints limit our training sample to 5,000 stars, balancing model performance with resource requirements.

### 3.4. Payne module

We implement the Payne algorithm (Ting et al. 2019) to predict flux variations across the 11D parameter space listed in Table 1. Our architecture in `PyTorch` features two hidden layers (40 neurons each) and an output layer with a size of 5400. These layers are connected with sigmoid activations for smooth interpolation between stellar parameters. The spectral library splits into training (70%) and validation (30%) sets, with mean squared error evaluated every 5,000 epochs during optimisation. We utilise the ADAM optimiser (Kingma & Ba 2014) with a learning rate of 0.001, monitoring validation loss to prevent overfitting.

### 3.5. Performance

We validate the spectroscopic model's performance on synthetic data prior to application to observations. A test set of 5,000 spectra — following the training label distribution but spanning SNR = 50–1000 — is generated independently of the training process. The full pipeline (SLAM + Payne modules, Sec. 3.1) is applied to evaluate three key metrics: (1) pixel-level spectral reconstruction (Appendix A), (2) label recovery accuracy across SNR regimes, and (3) consistency between repeated observations of the same target.

Fig. 1 shows the SLAM module's performance in recovering $T_{eff}$ and $\log g$ across the Kiel diagram. Results are binned by $\Delta T_{eff} = 1000$ K and $\Delta \log g = 0.5$ dex (gray dashed lines). Colored markers denote standard deviations for SNR = 30 (top), 50, 100, and 250 (bottom) within each populated bin. Two trends emerge: (1) precision improves systematically with SNR, particularly below SNR = 100, and (2) $T_{eff}$ accuracy degrades logarithmically with increasing temperature, from $\sigma_{T_{eff}} \approx 50$ K at 7000 K to $\sigma_{T_{eff}} \approx 300$ K at 17000 K. While $\log g$ precision remains relatively constant ($\sigma_{\log g} \approx 0.1$ dex) across most of the parameter space, tentative degradation occurs for $T_{eff} > 16000$ K giants ($\log g < 3.5$ dex), likely due to reduced line density in evolved stars.

In Fig. 2, we present the scatter (solid) and bias (dashed) of different labels as a function of SNR. The parameters $T_{eff}$, $\log g$, [Fe/H], and $v \sin i$ are estimated via the SLAM module, while the remaining labels are derived from the Payne module.

We categorize the test sample into three temperature ranges: 7000 K < $T_{eff}$ < 9000 K (blue), 9000 K < $T_{eff}$ < 13000 K

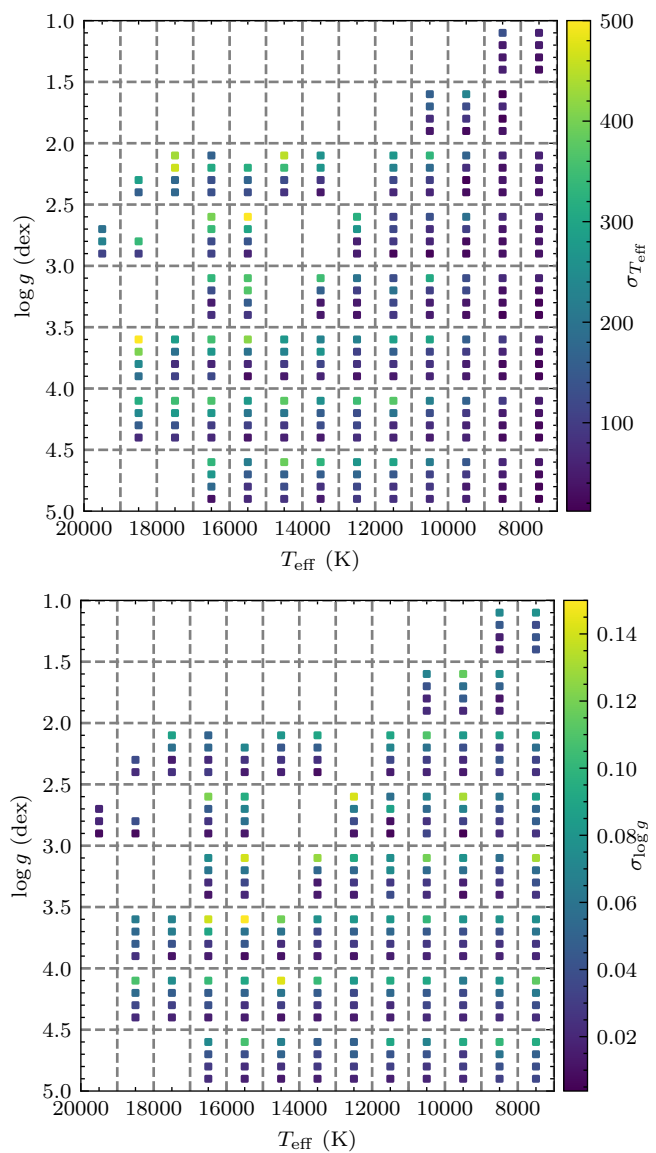

Fig. 1: Recovery precision for $T_{eff}$ (top) and $\log g$ (bottom) from the SLAM module across the Kiel diagram. Test spectra are binned by $\Delta T_{eff} = 1000$ K and $\Delta \log g = 0.5$ dex (delineated by gray dashed lines). For each bin with a significant number of spectra, the four "traffic light" bullets show the standard deviation of the label for SNRs of (top to bottom) 30, 50, 100, and 250. Empty bins contain insufficient data for robust statistics.

(green), and 13000 K < $T_{eff}$ < 17000 K (orange). A representative example at SNR = 100.0 (without $T_{eff}$ selection) is shown, with the bias ($\mu$) and scatter ($\sigma$) for each label marked in the upper-right corner of the corresponding panel. For nearly all labels, the most pronounced trend across the three subpopulations is the performance improvement (i.e., decreasing bias and scatter) with increasing SNR. This reflects the expected reduction in noise contamination at higher SNRs. However, [C/Fe], [S/Fe] and [Ca/Fe] exhibit a weaker dependence of $\sigma$ on SNR, with persistently large scatter ($\sigma > 0.35$). This indicates that, given the resolution and wavelength coverage of LAMOST LRS, the current model cannot reliably recover these abundances.

By comparing the behaviour of subsamples with different temperatures, we find that the remaining labels can be largely classified into two groups based on the degree of temperature dependency. For [O/Fe], [Mg/Fe], and [Si/Fe], different sub-samples show very similar profiles in their scatter and bias, with average scatters at SNR = 100 of 0.22, 0.14, and 0.17, respectively.

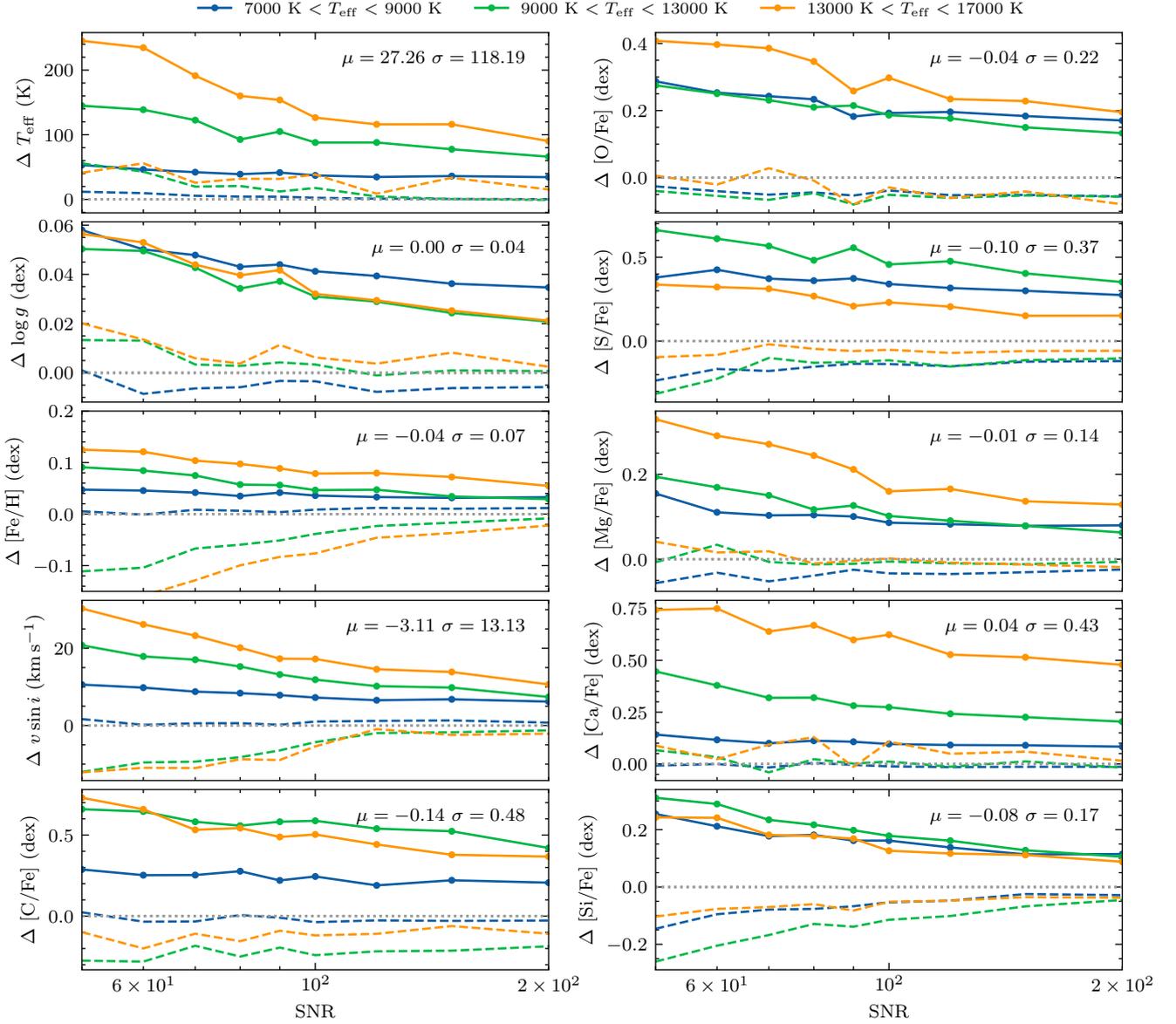

Fig. 2: Label recovery precision versus signal-to-noise ratio (SNR per Å) for the spectroscopic pipeline. Solid and dashed lines represent scatter ($\sigma$) and bias ($\mu$), respectively, calculated within SNR bins. Colored bands indicate three temperature regimes: $7000\,\mathrm{K} < T_{\mathrm{eff}} < 9000\,\mathrm{K}$ (blue), $9000\,\mathrm{K} < T_{\mathrm{eff}} < 13000\,\mathrm{K}$ (green), and $13000\,\mathrm{K} < T_{\mathrm{eff}} < 17000\,\mathrm{K}$ (orange). Insets show $\sigma$ and $\mu$ values at SNR = 100 in the units applicable to the respective labels for the full sample.

In contrast, the results for the other labels vary significantly among different temperatures. As stars become hotter, most spectral features available for measuring the labels weaken, which largely explains the trend observed in Fig. 2. Hotter stars generally exhibit larger scatter in $T_{\mathrm{eff}}$ compared to cooler stars. Microturbulence (not included in Fig. 2) displays similar behavior to $T_{\mathrm{eff}}$ and $v \sin i$, with a typical uncertainty of $\mu = -0.07\,\mathrm{km\,s^{-1}}$ and $\sigma = 0.58\,\mathrm{km\,s^{-1}}$ at SNR = 100. The variation of $\log g$ with temperature is not significant, as shown in Fig. 1. The dependency on $T_{\mathrm{eff}}$ is particularly important for [Fe/H], where the absolute value of bias significantly increases with higher temperatures. This again illustrates the difficulty of inferring abundance for hotter stars using low-resolution spectra and hints at a possible missing population of hot, metal-poor stars. However, we do not observe such a strong dependence in bias for other labels of [X/Fe], which may be due to the relative ratios between spectral lines of a given abundance and those of iron lines.

While these tests demonstrate the accuracy of the model reconstruction and establish the minimum uncertainty for the recovered parameters, larger uncertainties could be expected for observed data. This is due to various factors, such as the physical complexity of model atmospheres, synthetic spectra, and characteristics of the observed data (including data reduction effects), which contribute to the overall uncertainties. We then use targets with repeated observations and take the scatter of the labels inferred from these observations as the precision of the Payne model.

Fig. 3 presents the scatter of labels between repeated observations. This is based on 36,563 spectra of 2,034 stars, with a median observation number of nine. The $SNR_r$ is defined as the average value of the SNR in the $r$ band, and the scatter is calculated as the standard deviation of the labels among repeated observations for a given star, weighted by their SNRs. The median value of the scatter in bins of size 50 in SNR is shown in orange, demonstrating a photon noise-dominated phase for low SNR. We adopt the median value of the scatter as the precision of our sample, which is 163 K for $T_{\mathrm{eff}}$, 0.12 dex for $\log g$, 0.10 dex for [Fe/H], 18.8 km s$^{-1}$ for $v \sin i$, 0.5 km s$^{-1}$ for $v_{\mathrm{mic}}$ (not included in Fig. 3) and approximately 0.2 dex for [X/Fe].

This is roughly consistent with the estimation in Fig. 2 within a factor of 2, despite the differences in the samples between the two tests.

Given the large uncertainties of [C/Fe], [S/Fe] and [Ca/Fe], we have decided to exclude C, S, and Ca from subsequent analysis. Therefore, aside from iron abundance, we will present the results for the three $\alpha$-elements ([O/Fe], [Mg/Fe], and [Si/Fe]) for the sample.

## 4. Method validation

In this section, we present verification and comparison results with literature studies. We separate these into two parts: verification against high-resolution data (Sec. 4.1) and comparison against low-resolution data (Sec. 4.2). We define high resolution as spectral resolutions exceeding $R = 10000$.

### 4.1. Verification against high-resolution data

Given the brightness limitations of the LAMOST LRS, relatively few reference stars have well-established parameters and abundances derived from high-resolution NLTE spectroscopy. We therefore combine reference stars with published NLTE/LTE analyses and synthetic spectra generated from high-resolution observations.

Our verification sample incorporates benchmark stars from Przybilla et al. (2006), Takeda et al. (2008), and Gebruers et al. (2021). Using FEROS echelle spectra ($R \sim 48,000$) from the MPG/ESO 2.2-meter telescope, Przybilla et al. (2006) developed a hybrid NLTE spectrum synthesis technique to analyse four BA-type supergiants, deriving abundances for He, C, N, O, Mg, S, Ti, and Fe. Takeda et al. (2008) obtained $R \sim 45,000$ spectra with the Bohyunsan Observatory Echelle Spectrograph on the 1.8 m reflector, measuring abundances of seven elements in 46 bright A-type stars while considering NLTE effects exclusively for the O I triplet at 7771–7775 Å. Gebruers et al. (2021) analysed $R \sim 85000$ HERMES spectra from the 1.2 m Mercator telescope for 111 pulsating stars, determining atmospheric parameters via the *Grid Search in Stellar Parameters* method (Tkachenko 2015) under LTE assumptions. Fig. 4 displays the benchmark stars' temperature-gravity distribution (top) and chemical abundances (bottom). The sample spans effective temperatures of 8000-19000 K and iron abundances from $\sim -1.0$ dex to $\sim +0.5$ dex, including three giants ($\log g < 3.5$). We identify tentative correlations between [O/Fe] and [Fe/H], as well as between [Si/Fe] and [Fe/H], though larger samples are needed to confirm these trends.

The verification results appear in Fig. 5, showing differences between our derived labels and literature values as functions of effective temperature. For stars from Przybilla et al. (2006) and Takeda et al. (2008) that are too bright for LAMOST-LRS, we retrieve archival XSHOOTER/UVES spectra degraded to LAMOST resolution (open squares). Literature [X/H] values were calculated from published [X/H] and [Fe/H] measurements.

Our results show good consistency for $T_{\mathrm{eff}}$, $\log g$, and $v \sin i$ without systematic trends with temperature. One supergiant ($\log g = 1.20$, not shown in Fig. 5) exhibits significant off-sets: $\Delta \log g = -0.6$ dex and $\Delta v \sin i = +150$ km s$^{-1}$, likely from NLTE effects influencing both gravity and rotation measurements.

Iron abundances generally match literature values with temperature-independent scatter ($\sigma_{\mathrm{[Fe/H]}} \sim 0.2$ dex). While deviations increase slightly for $T_{\mathrm{eff}} > 15000$ K, this coincides with larger literature uncertainties. Hot metal-poor stars ([Fe/H] < $-0.5$ dex, $T_{\mathrm{eff}} > 15000$ K) may show underestimated metallicities, consistent with the negative bias for hot stars in Fig. 2. Light elements demonstrate varying performance: [Mg/Fe] and [Si/Fe] show minimal temperature dependence with $\sigma \sim 0.3$ dex, though the hottest stars exhibit slight underestimates. This could arise from literature [Fe/H] overestimates affecting abundance ratios. Oxygen abundances display larger scatter ($\sigma_{\mathrm{[O/Fe]}} \sim 0.4$ dex) without clear temperature trends.

Our verification demonstrates good agreement with high-resolution studies across $T_{\mathrm{eff}}$ (7500-20000 K) for fundamental parameters and key abundances. Scatter remains within 0.2 dex for [Fe/H], 0.3 dex for [Mg/Fe] and [Si/Fe], and 0.4 dex for [O/Fe], comparable to literature uncertainties. Increased dispersion at the hot end reflects intrinsic measurement challenges and potential systematic differences between analysis methods.

### 4.2. Comparison against low-resolution data

While high-resolution spectroscopy provides precise benchmarks, comparison with low-resolution surveys remains essential for validating pipeline performance across large samples. The statistical power of LRS datasets enables the detection of population-level trends obscured in smaller high-resolution samples, though reduced spectral resolution requires careful systematic error assessment.

We compare our results with two major studies: Xiang et al. (2022) using LAMOST LRS ($R \sim 1800$) with the Hot-Payne method, and Sun et al. (2021a) analyzing LAMOST MRS ($R \sim 7500$) spectra covering 4950–5350 Å and 6300–6800 Å. For robust comparison, we restrict the LRS-based sample to stars with $SNR_r > 100$. The cross-matched sample with Xiang et al. (2022) consists of 65,590 stars, while the overlap with the MRS-based analysis includes 8,078 stars, spanning $7000 - 14500$ K with median $SNR_B \sim 40$. Additionally, 10,713 stars are common to both external catalogues.

Fig. 6 compares four key parameters across studies. Effective temperatures (Panel a) show good agreement ($\sigma = 442$ K) between this work and Xiang et al. (2022), across the entire temperature range. The MRS comparison reveals a ~5% coolward bias below 12000 K, potentially from differences in wavelength coverage between LRS and MRS, i.e., MRS only covers the H$\alpha$ line. Enhanced scatter around A0-type stars (9000–11000 K) suggests challenges in modelling Balmer line morphology transitions in LRS data (see Fig. 7). This may also explain the difference of scatter when we cross-match the results to Sun et al. (2021a), where the top-left panel shows a tighter correlation compared to the bottom-right one.

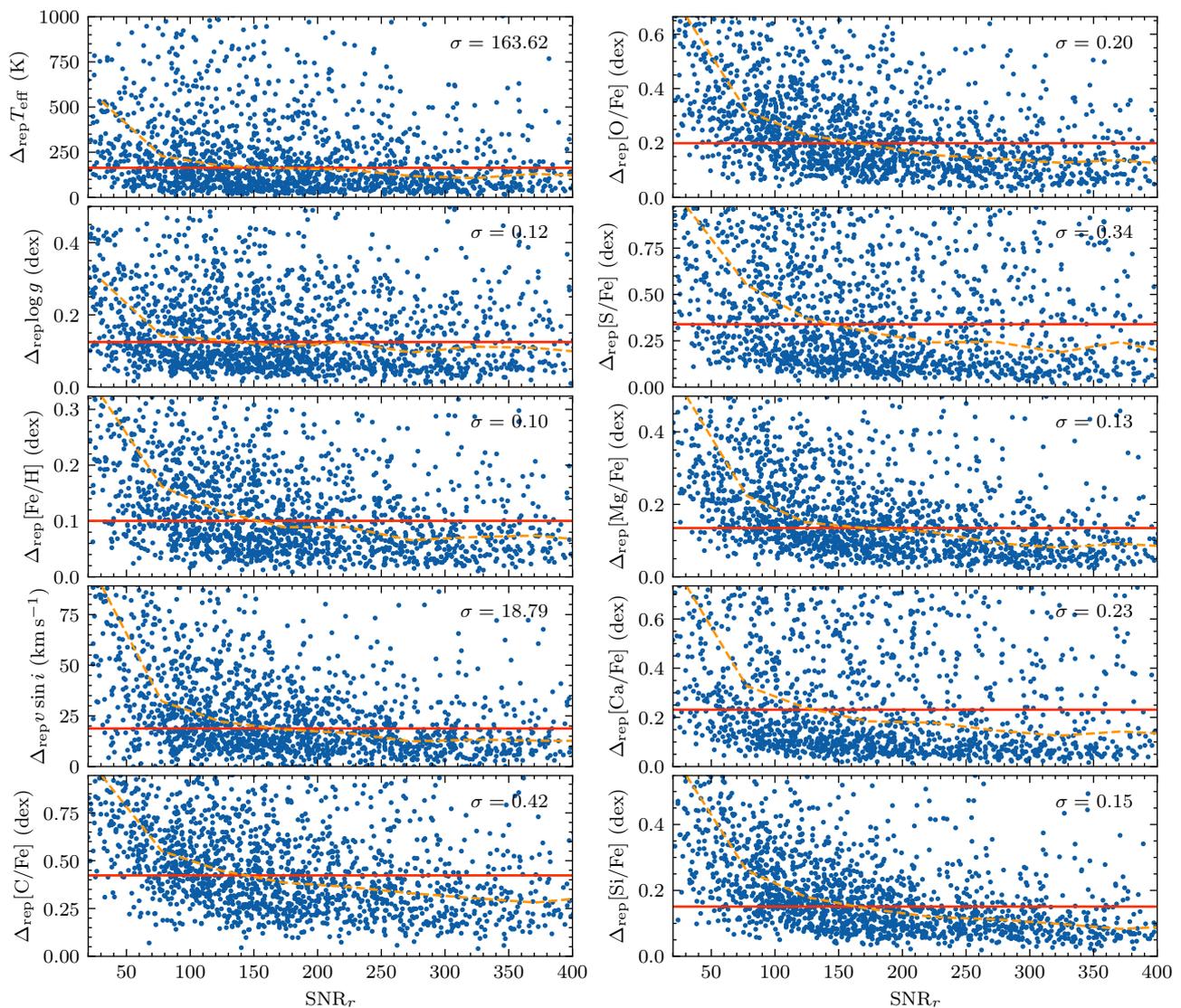

Fig. 3: Scatter of labels between repeated observations. The $SNR_r$ is defined as the average value of the SNR in the $r$ band of the observations, and the scatter $\Delta_{rep}l$ is the standard deviation of the labels in the units applicable to the respective labels, weighted by their SNR. The orange dashed lines represent the median value of the scatter in each SNR bin. The red horizontal lines indicate the median value of $\Delta_{rep}l$ from the entire sample with repeated observations, which are also marked in the top-right corner.

As for surface gravity comparisons (Panel b), we focus on Xiang et al. (2022) since MRS spectra provide limited $\log g$ constraints due to their narrow wavelength coverage (Zhang et al. 2020b). Our $\log g$ measurements show a systematic offset of 0.11 dex relative to Xiang et al. (2022), predominantly driven by stars in the 9000–11000 K range. The white contours in Fig. 6 explicitly demonstrate how this temperature-dependent bias distorts the correlation. We attribute this discrepancy to differences in gravity-sensitive feature selection between pipelines, particularly for A-type stars, where Balmer line profiles dominate LRS spectra.

Metallicity comparisons (Panel c) reveal good overall agreement with Xiang et al. (2022) ($\sigma_{[Fe/H]} \sim 0.21$ dex). We identify a −0.14 dex systematic offset for stars near solar metallicity ([Fe/H] ≈ 0), which diminishes at both metal-rich ([Fe/H] > +0.3) and metal-poor ([Fe/H] < −0.5) extremes. This pattern persists across the 9000–11000 K subsample (white contours), suggesting temperature-independent calibration differences rather than physical abundance variations. The bias when compared against Sun et al. (2021a) could be attributed to the difference between [Fe/H] and [M/H] .

The last panel presents the comparison for $v \sin i$. This work and Xiang et al. (2022), both based on LAMOST-LRS, are consistent with each other and largely follow the same profile when compared with Sun et al. (2021a). In particular, rotational velocities derived by Sun et al. (2021a) are larger than those derived in this work for $v \sin i \leqslant 100$ km s$^{-1}$ and are in good agreement for larger values. This behaviour is expected due to the medium-resolution data used in Sun et al. (2021a).

To investigate label differences across stellar spectral types, Fig. 7 displays label variations as a function of effective temperature. The subplots, from top to bottom, show the differences $\Delta l = l_{Xiang} - l$ for the four atmospheric parameters $T_{eff}$, $\log g$, [Fe/H] , and $v \sin i$ derived from the SLAM module. Similar to Fig. 6, $T_{eff}$ values show generally good agreement, though the difference distribution becomes asymmetric about $\Delta l = 0$ (dashed line) for stars between 9500 K ($\log T_{eff} \sim 3.95$) and 11000 K ($\log T_{eff} \sim 4.05$). A comparable asymmetry appears in the log $g$ panel (second row). As noted previously, this likely stems from the HotPayne approach employed by Xiang et al. (2022), which masks Balmer line regions when analyzing A0-type stars — a simplification that degrades label recovery performance, particularly for $T_{\rm eff}$ and log $g$.

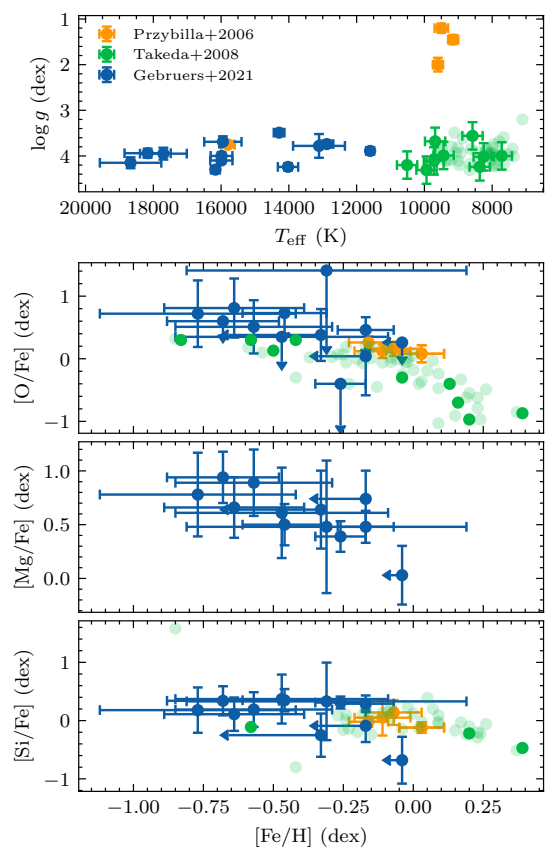

Fig. 4: Kiel diagram (top panel) and abundance patterns (bottom panel) for benchmark stars. Symbols denote different literature sources: Przybilla et al. (2006, orange), Takeda et al. (2008, green), and Gebruers et al. (2021, blue). Faded markers indicate stars lacking available spectra for direct verification.

The [Fe/H] panel (third row) reveals a systematic offset of $-0.13$ dex, most pronounced below $T_{\rm eff} < 10000$ K. This metallicity bias gradually diminishes with increasing temperature, culminating in a steep transition near the grid boundary at $T_{\rm eff} \sim$ 19000 K (log $T_{\rm eff} \sim 4.25$). The $v \sin i$ panel (bottom row) features a temperature-independent horizontal line at $-5\,{\rm km\,s^{-1}}$, reflecting differing lower limits in rotational velocity between studies. Our analysis adopts a $5\,{\rm km\,s^{-1}}$ minimum for projected rotational velocity, compared to the $0\,{\rm km\,s^{-1}}$ threshold used by Xiang et al. (2022). This discrepancy produces an artificial offset population corresponding to slow rotators present in both samples.

We further analyse the $v \sin i$ distribution by comparing our results with two other studies, focusing on low rotational velocities. Accurate determination of projected rotational velocity from low-resolution spectra (LAMOST-LRS) is challenging at small $v \sin i$ values, as rotational broadening blends with instrumental broadening. Fig. 8 shows the $v \sin i$ distributions and complementary cumulative distribution functions. Results from both LAMOST-LRS studies agree well with LAMOST-MRS measurements for $v \sin i \geqslant 200\,{\rm km\,s^{-1}}$. However, we note an anomalously high fraction of slow rotators ($v \sin i \leqslant 20\,{\rm km\,s^{-1}}$) in Xiang et al. (2022), despite these velocities being below typ-ical detection limits. Discrepancies emerge at lower velocities: For $v \sin i = 100 - 200\,{\rm km\,s^{-1}}$, Xiang et al. (2022) shows a 10–20% deficit compared to Sun et al. (2021a), while our results differ by $< 5\%$ from the latter. This trend appears more clearly in the cumulative distribution (bottom panel), where the distributions begin to diverge below $100 - 120\,{\rm km\,s^{-1}}$. Previous spectral resolution studies (Sun et al. 2019, 2021b) estimate a $120\,{\rm km\,s^{-1}}$ detection limit for LAMOST-LRS (gray dashed line in Fig. 8), consistent with the velocity range where significant discrepancies occur. This suggests improved $v \sin i$ accuracy in our work for $v \sin i > 120\,{\rm km\,s^{-1}}$, though results below this threshold should be interpreted cautiously.

These comparisons reveal how spectral resolution and methodology affect parameter determination. While high-resolution data provide precision benchmarks, the consistency between low-resolution studies demonstrates our pipeline's robustness for large samples. Systematic biases in log $g$ and [Fe/H] for A0 stars, however, highlight limitations in both the HotPayne method and data processing approaches when critical spectral features are masked.

## 5. Results

In this paper, we employ the SLAM method in conjunction with the Payne model to produce a catalogue of stellar properties for early-type stars derived from LAMOST LRS DR9. This section outlines our sample filtering criteria for the final catalogue and provides examples of how our catalogue can investigate the early-type stellar population within the survey's footprint, including the blue horizontal branch stars, abundance distribution, and abundance gradient.

### 5.1. Sample filtering

We implement a selection criterion based on the SNR in the $r$ band, excluding spectra with SNR lower than 10. As illustrated in Fig. A.2, the spectral information for early-type stars is limited, particularly for those with higher effective temperatures. To ensure the reliability of our results across different temperature ranges, we recommend a more conservative SNR threshold of 50, especially for users interested in abundance determinations. Additionally, we apply several cuts based on the Hess diagram of the remaining candidates. First, we discard stars with $T_{\rm eff} \leqslant 7100$ K, as those near the lower boundary are likely to be cooler stars with $T_{\rm eff} \leqslant 7000$ K. This exclusion removes more than half of the candidates from the sample. Similarly, we impose a cut at 19000 K to eliminate stars with potentially higher temperatures. These temperature cutoffs are informed by comparisons with existing studies, specifically Luo et al. (2015); Sun et al. (2021a); Xiang et al. (2022).

We present the Kiel diagram of our sample in Fig. 9, overlaid with MIST evolutionary tracks for stars ranging from 1.7 to $7.0\,{\rm M_\odot}$. An abnormal feature is evident in the top-left corner of the Kiel diagram, extending from the hot dwarf region toward cooler giants. By comparing this with catalogs extending to higher temperatures, we confirm that these features are artifacts arising from hotter stars beyond the limitations in the temperature grid ($T_{\rm eff} = 20000$ K). Consequently, we exclude all candidate stars above the dashed line. Ultimately, the final sample consists of 315,822 stars with SNR $\geqslant 10$ and 195,004 stars with SNR $\geqslant 50$. In contrast to the results presented in Xiang et al. (2022) (see their Fig. 13), we do not observe a vertical stripe of stars extending to low log $g$ values. As noted by Xiang et al. (2022), this feature is an artefact caused by a large number of chemically peculiar stars, whose $\log g$ values may have been significantly underestimated due to the masking of hydrogen lines during label determination for early A-type and late B-type stars.

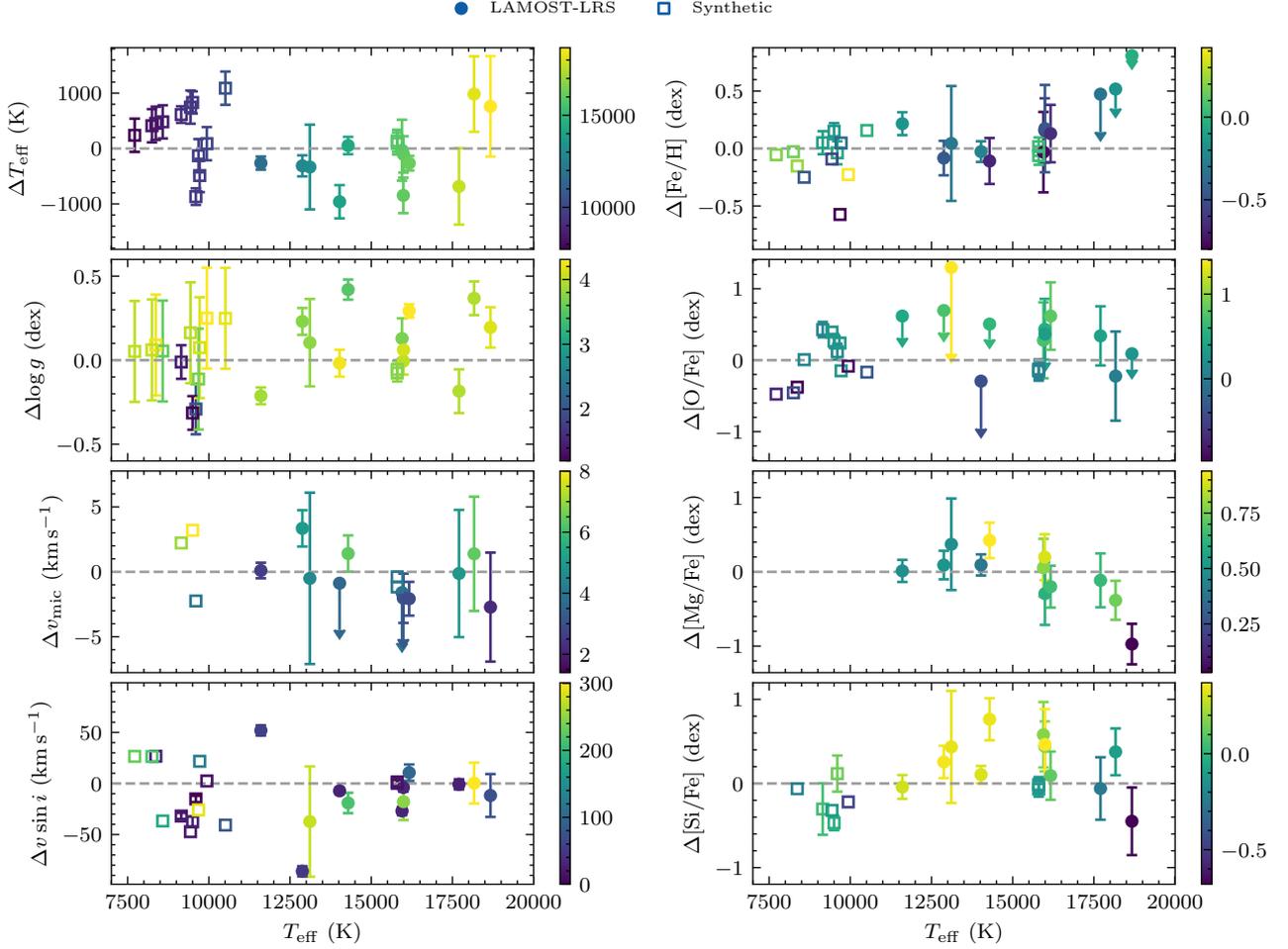

Fig. 5: Label differences $\Delta l = l_{lit} - l$ between literature results from high-resolution spectroscopy and this work, plotted against $T_{eff}$. Colored symbols denote benchmark stars with high-resolution data, where colour intensity corresponds to label values. Circles indicate stars with LAMOST LRS spectra, while empty squares represent spectra mimicked from high-resolution observations. Error bars and upper limits reflect uncertainties from Przybilla et al. (2006), Takeda et al. (2008), and Gebruers et al. (2021).

## 5.2. Photometric selection

The preselection (Sec. 2) and data cleaning (Sec. 5.1) procedures for our sample are primarily driven by spectroscopic criteria. While a selection based on $\tilde{M}_{K_s}$ is implemented to eliminate contamination from cooler stars, this approach essentially functions as a magnitude-cutting method that does not utilise colour information, which could enhance sample purity. Therefore, the remaining stars can be used to verify the effectiveness of the photometric selection criteria. Zari et al. (2021) combined *Gaia* EDR3 astrometry and photometry with 2MASS photometry to create an all-sky sample of luminous OBA-type stars. They employed two sets of selection rules based on the colour-colour diagram and colour-magnitude diagram, defined by the following equations:

$$J - H < 0.15(G - K_s) + 0.05$$
$$J - H > 0.15(G - K_s) - 0.15 \tag{1}$$

and

$$G > 2(G - K_s) + 3 \tag{2}$$

The first set of criteria selects O- and B-type stars along a sequence in the $G - K_s$ versus $J - H$ colour-colour diagram, accounting for interstellar reddening, which effectively separates these stars from redder turn-off stars and giants. The second set further distinguishes cool giants from OBA stars.

In Fig. 10, we present the distribution of the sample with SNR $\geqslant 50$ in the $J - H$ vs. $G - K_s$ colour-colour diagram (left) and the $G$ vs. $G - K_s$ colour-magnitude diagram (right). The photometric selection criteria for OBA stars, as defined by Zari et al. (2021), are indicated by red dashed lines and the grey shaded area. Stellar density is represented on a logarithmic scale, with black contours showing the distribution of the entire sample at kernel density levels of 25%, 50%, and 75%.

Over 95% of our sample falls within the region defined for OBA stars by Zari et al. (2021). However, some outliers may be contaminants from redder turn-off stars and giants. For instance, Zari et al. (2021) attributed the outliers above the red dashed line in the right panel to residual giants. Our estimated stellar parameters for these candidates are primarily clustered near the lower limit of the temperature grid, indicating that they are indeed contaminants.

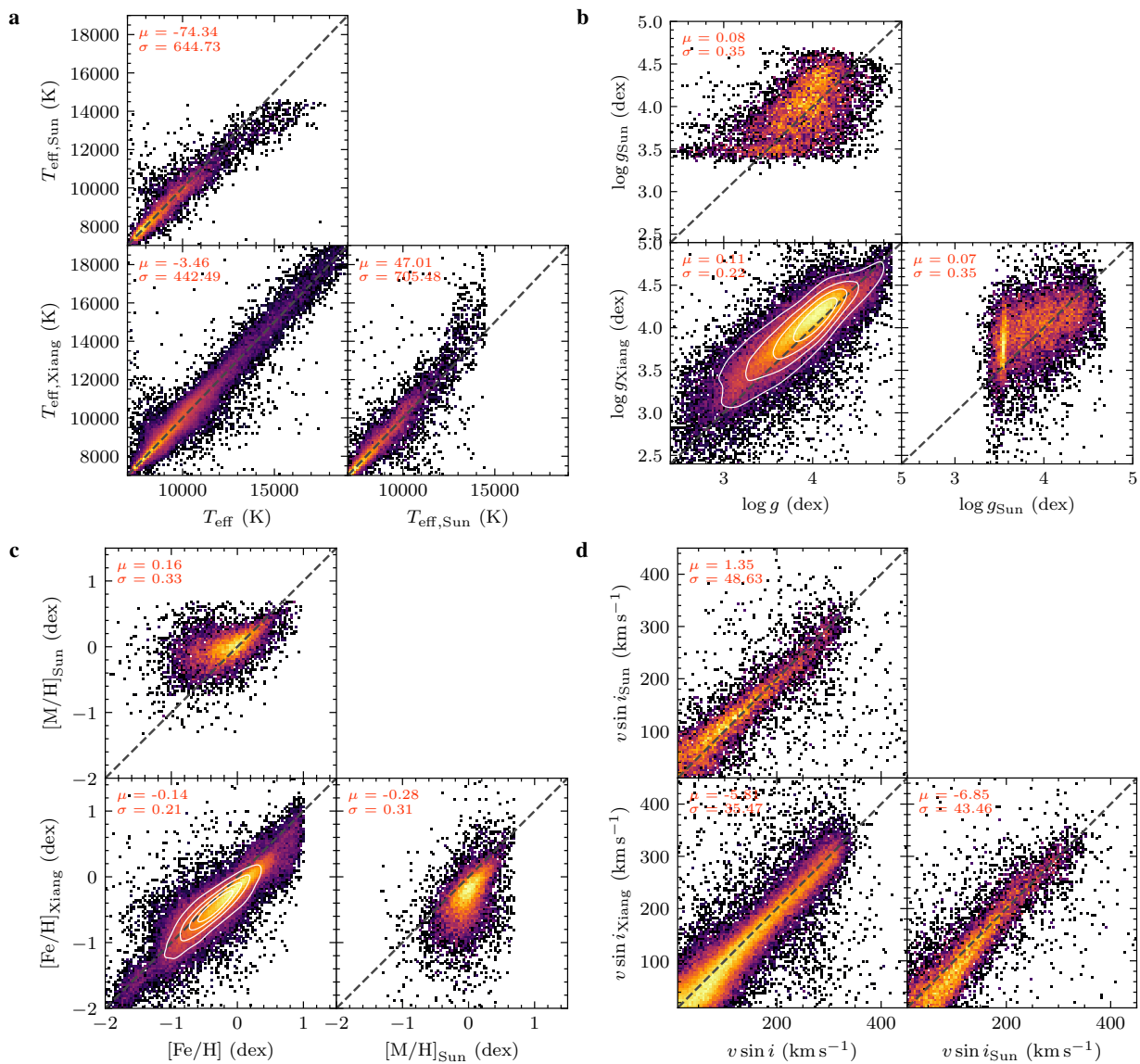

Fig. 6: Parameter comparisons with Xiang et al. (2022, LRS) and Sun et al. (2021a, MRS). Colour density scales with log($N$). Panels (b)–(c) overlay contours (white) for 9000–11000 K stars. Dashed lines show 1:1 relations, with red text indicating mean offsets ($\mu$) and standard deviations ($\sigma$).

Other than this, we note that in the left panel, stars exhibit a broader spread in $G-K_s$, while the majority are distributed along a shaded region. Although we cannot definitively rule out residual contamination in our sample, these outliers are likely intrinsic hot stars, as their distributions in the Kiel diagram and $T_{\rm eff}$ versus [Fe/H] space show no anomalous clustering. A similar pattern is evident in Fig. 1 of Zhang et al. (2020b) for the LAMOST OB stars from Liu et al. (2019), whose sample was selected based on distributions in spectral line indices (H$\gamma$, Ca II K, He I, and Fe). This highlights potential systematic differences between spectroscopic and photometric selection methods, as discussed in Zari et al. (2021) regarding sample completeness. For example, approximately 9% of the stars in the Liu et al. (2019) catalogue do not satisfy their photometric selection criteria, indicating that approximately 5% to 10% of hot star candidates may be missed.

### 5.3. Blue horizontal branch stars

Blue horizontal branch stars are old, metal-poor Population II stars with masses less than $1.0\,{\rm M}_\odot$, typically found in the Galactic halo (Kinman et al. 1994). They are core helium-burning stars located on the blue side of RR Lyrae variables in the Hertzsprung-Russell diagram, characterised by strong hydrogen lines and weak or no molecular bands in their spectra. Their nearly constant absolute magnitude within a restricted colour range makes them useful as distance tracers in studying the Milky Way's structure and kinematics (e.g., Pier 1984).

In this work, we identify BHB candidates using Galactic latitude ($|b|$) and iron abundance criteria ([Fe/H] ). Fig. 11 reveals that the metallicity distribution of stars with SNR$_r \geqslant 50$ contains a low-metallicity tail extending from [Fe/H] $\sim -1.0$ dex. Most metal-poor stars in this tail display significantly smaller projected rotational velocities compared to the main population, approaching the detection limits imposed by the spectral resolution (see Sec. 4.2). These metal-poor stars are preferen-tially located at greater heights above the Galactic plane, consistent with expectations for BHB stars and blue stragglers. Another subpopulation shows moderate $v \sin i$ and high metallicity ([Fe/H] > 0.5 dex), likely corresponding to chemically peculiar stars.

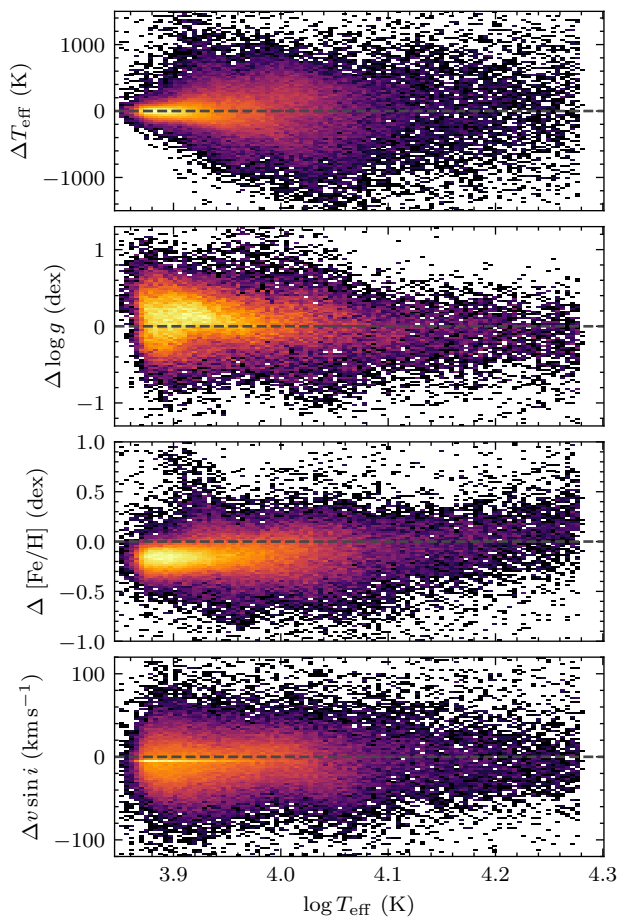

Fig. 7: Difference $\Delta l = l_{Xiang} - l$ as a function of $T_{eff}$ from the SLAM module. Subplots from top to bottom show results for $T_{eff}$, $\log g$, [Fe/H], and $v \sin i$ respectively.

Fig. 12 shows the distribution of BHB candidates in the $T_{eff}$–$\log g$ plane, color-coded by their projected rotational velocities. Our sample selects stars with Galactic latitude $|b| > 15°$, metallicity [Fe/H] < −1.4 dex, and spectral signal-to-noise ratio $SNR_r \geq 50$, yielding 3564 candidates. Two distinct populations emerge: over 90% of candidates have $\log g < 3.5$ with $T_{eff}$ reaching ∼ 11000 K, while the remaining 10% reside closer to the main sequence.

We overlay PARSEC isochrones (Bressan et al. 2012) for core-helium-burning horizontal branch stars with metallicities [Fe/H] = −2.5 dex (solid) and −2.0 dex (dashed), computed for ages $\log(t/yr)$ = 8.6, 8.8, and 9.0. The isochrones show excellent agreement with stellar parameters for candidates having $\log g < 3.5$ dex, confirming their identification as genuine BHB stars.

The higher surface gravity population ($\log g \geq 3.5$ dex) likely consists of blue straggler stars (BSS). Notably, BSSs exhibit systematically higher rotational velocities than BHB stars near the horizontal branch. This aligns with expectations that BSS rotation slows through angular momentum loss mechanisms like magnetic braking and disk locking (Leonard & Livio

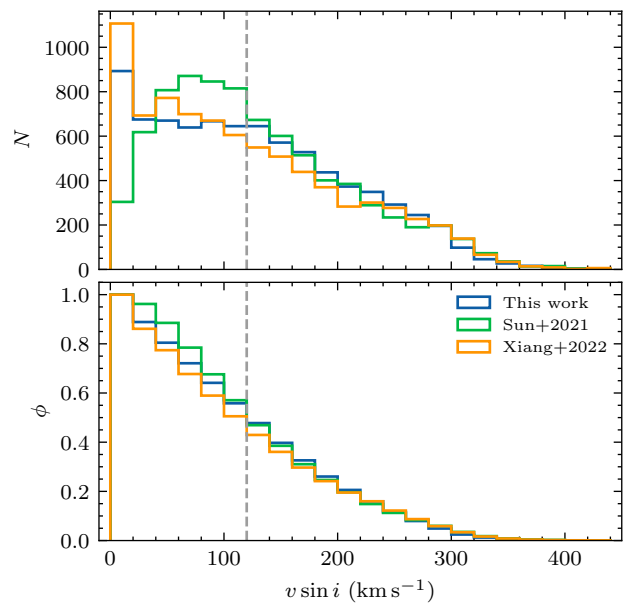

Fig. 8: Distribution ($N$, top) and complementary cumulative distribution functions ($\phi$, bottom) of $v \sin i$ for cross-matched samples in this work (blue), Sun et al. (2021a) (green), and Xiang et al. (2022) (orange). The gray dashed line at $v \sin i = 120 \, \text{km s}^{-1}$ marks the estimated lower threshold for reliable $v \sin i$ determination at LAMOST-LRS resolution.

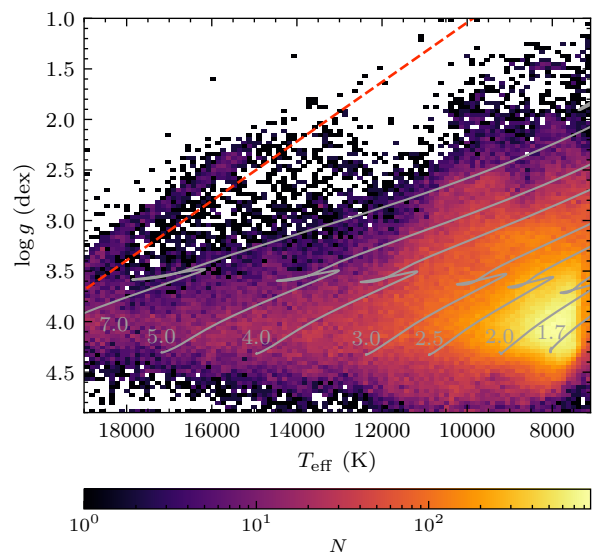

Fig. 9: Distribution of candidate stars with $r$-band SNR > 10 in the Kiel diagram. Stellar density is colour-coded on a logarithmic scale. Temperature cuts at 7100 K and 19000 K have been applied. Stars above the red dashed line represent artifacts from intrinsically hotter stars ($T_{eff} \geq 20000$ K) and are excluded from further analysis. Grey curves show evolutionary tracks for stars with masses of 1.7, 2.0, 2.5, 3.0, 4.0, 5.0, and 7.0 M$_\odot$ from MIST (Paxton et al. 2011; Choi et al. 2016) models.

1995; Leiner et al. 2018). While BHB stars rotate more slowly than young disk stars due to their advanced age, our results suggest they may rotate slower than BSS at comparable metallicities. This is consistent with the scale-width-shape method used in Clewley et al. (2002) and Xue et al. (2008) to differentiate BSS from BHB stars, based on the shapes of the Balmer lines. A small population of horizontal branch stars (~10) exhibit anomalously high rotation ($v \sin i > 150 \, \text{km s}^{-1}$). These outliers likely represent high-luminosity BSS contaminating the horizontal branch region, possibly originating from evolved dwarf stars.

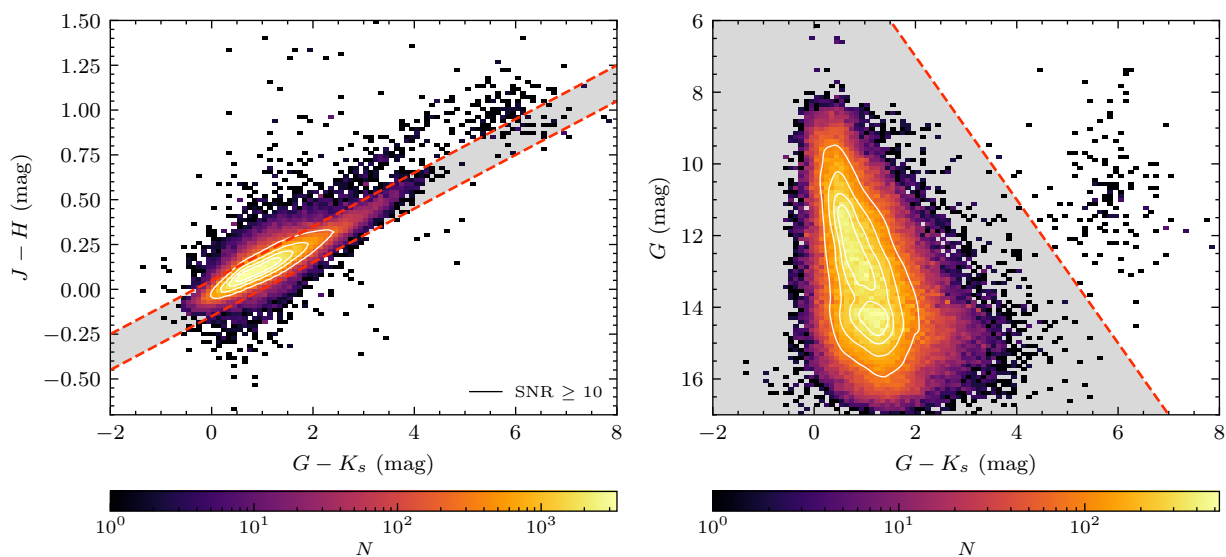

Fig. 10: $J - H$ vs. $G - K_s$ color–color diagram (left) and $G$ vs. $G - K_s$ color-magnitude diagram (right) for the sample with SNR $\geqslant 50$. Stellar density is colour-coded on a logarithmic scale. Density contours derived from kernel density estimates are white for the entire sample. The red dashed lines and grey shaded area delineate the selection criteria for OBA stars as outlined by Zari et al. (2021).

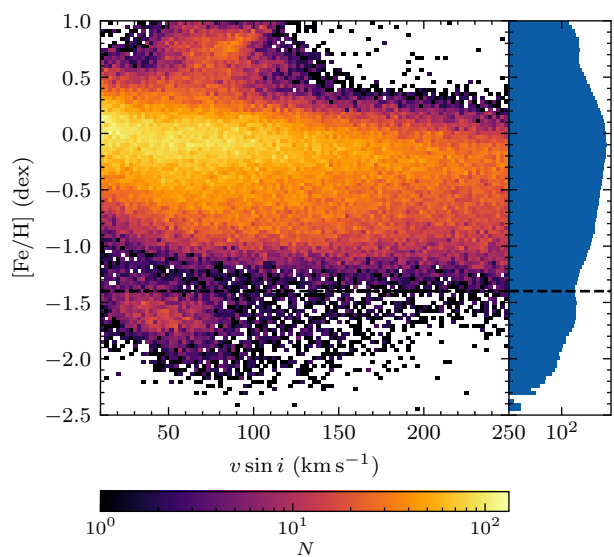

Fig. 11: Distribution of stars with $SNR_r \geqslant 50$ in [Fe/H] and $v \sin i$ space. The right subplot displays the logarithmic histogram distribution of [Fe/H], with a horizontal dashed line indicating the metallicity threshold ([Fe/H] $= -1.4$ dex) for BHB candidate selection.

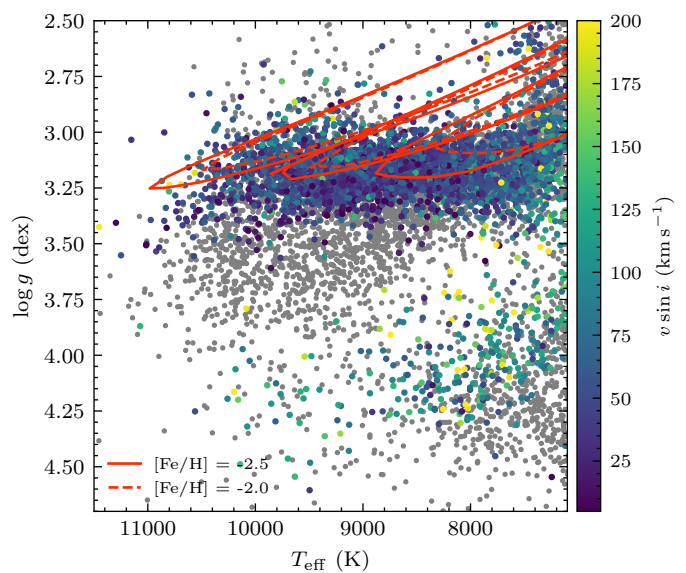

Fig. 12: 3564 BHB candidates in the Kiel diagram, color-coded by projected rotational velocities. Grey dots show BHB candidates selected by Xiang et al. (2022) using similar selection criteria. Red lines display PARSEC (Bressan et al. 2012) isochrones for horizontal branch stars with metallicities [Fe/H] $= -2.5$ dex (solid) and $-2.0$ dex (dashed), corresponding to ages $\log(t/\text{yr}) = 8.6$, $8.8$, and $9.0$ (left to right respectively).

In Fig. 12, we also show the BHB candidates selected from Xiang et al. (2022) as gray points, following the same selection criteria as those applied in this work (with the sole difference being our use of SNR in the $g$ band rather than the $r$ band). We find good agreement in the stellar labels for stars with $T_{\text{eff}} \sim 8000$ K and $\log g \sim 3.25$ dex. However, the labels from

Xiang et al. (2022) exhibit significantly higher $\log g$ values for BHB stars hotter than this temperature. This discrepancy, as discussed in Sec. 4.2, may arise from the HotPayne method employed in their work, which can lose spectroscopic information near Balmer lines for A-type stars. Further exploration of the completeness of the BHB sample and its usage as distance indicators will be studied in a forthcoming paper (Sun et al., in prep.).

## 5.4. Abundance distribution

Fig. 13 shows the stellar density distributions in abundance space for the sample with $SNR_r > 100$. We separate the sample into two temperature ranges: cooler than 10000 K and hotter than 10000 K. This division reflects the intrinsically different abundance patterns observed between these groups. The verification sample from Sec. 4.1 is shown as colored circles with error bars, matching the colour scheme of Fig. 4.

The abundance trends align with literature results, particularly for stars cooler than 10000 K (left panel). However, the [Mg/Fe] and [O/Fe] abundances for hotter benchmark stars lie near the upper envelope of the derived pattern, with some literature values corresponding to the upper limits of chemical abundance measurements. Perfect agreement is not expected due to differences in selection functions between our sample and the literature samples.

We overplot the $\alpha$ elements abundances versus [Fe/H] for a sample of 180 individual Classical Cepheids from Trentin et al. (2024). These abundances were derived from high-resolution spectroscopy, covering a metallicity range from -1.0 dex to 0.25 dex. The temperature-based separation does not naturally apply to Cepheids; therefore, these stars, primarily serving as a reference, remain the same in both columns. Both early-type stars and Cepheids are young populations that form from the same well-mixed interstellar medium, so their initial metallicity and abundance distributions are expected to be similar. Although early-type stars require complex non-LTE corrections due to their high temperatures and Cepheids have phase-dependent atmospheric variations because of pulsation, once these effects are properly corrected, the derived abundances converge to reflect the same underlying chemical composition of the current interstellar medium. Stars with $T_{\text{eff}} < 10000$ K generally follow the reference trend, despite an offset in [O/Fe] being observed between our sample and the Cepheids, and also between the benchmark stars and the Chpheids. In contrast, hotter stars exhibit a distinct abundance pattern, characterized by a steeper slope between for [Fe/H] between $-0.5$ and 0.0 dex compared to cooler stars.

Measuring oxygen abundances is particularly challenging because the few accessible oxygen lines are either weak or blended — such as the [O I] 6300 Å line, which is contaminated by a nearby Ni I blend (Caffau et al. 2013) — or they are strongly affected by NLTE effects that require complex, model-dependent corrections (Takeda & Honda 2016). In early-type stars, high temperatures intensify NLTE effects and complicate line formation, whereas in Cepheids the dynamic, pulsating atmospheres introduce additional phase-dependent uncertainties. These differences contribute to systematic offsets in the derived [O/Fe] ratios between the two populations (Nissen et al. 2014) (see also in Fig. A.2).

As discussed in Sec. 5.3, the same super metal-rich ([Fe/H] > 0.5 dex) population appears in both the cool and hot subsamples, with [O/Fe] showing a significantly narrower scatter compared to the other two abundances. This subpopulation is likely an artefact caused by chemical peculiar stars whose chemical abundances are not well recovered, as our methodology may not be capable of disentangling abundance information for these stars — particularly since [Fe/H] and [X/Fe] are derived independently through separate models.

In addition to the super metal-rich population, a potential low-$\alpha$ sequence is visible in the [O/Fe]–[Fe/H] and [Mg/Fe]–[Fe/H] planes. For example, a bifurcation emerges in the [Mg/Fe] subplots at [Mg/Fe] < −0.2 dex. This feature persists across both temperature regimes at approximately the same abundance level. Notably, several verification stars from Takeda et al. (2008) occupy this low-$\alpha$ sequence, with [O/Fe] < −0.4 dex (top-left panel of Fig. 13). While less distinct in the [Si/Fe]–[Fe/H] plane (bottom-left panel of Fig. 13), a subtle signature of this sequence may exist in the hotter subsample. The origin and significance of this feature remain unclear at present, warranting further investigation to determine whether it arises from astrophysical processes or systematic effects.

## 5.5. Abundance gradient

The metallicity gradient in the Milky Way disk is a key observational constraint on models of Galactic chemical evolution and disk formation (Chiappini et al. 1997). Studies using a range of tracers — from classical Cepheids to H II regions, open clusters, and field red giants — consistently reveal a negative gradient of roughly −0.04 to −0.06 dex kpc$^{-1}$ (Genovali et al. 2014, and references therein). Classical Cepheids have been instrumental because their well-calibrated period–luminosity and period–metallicity relations yield robust measurements of the present-day gradient (Genovali et al. 2014; Trentin et al. 2023).

In the context of Galactic evolution, such a negative gradient is expected from an inside-out formation scenario where the inner regions experience more rapid star formation and chemical enrichment. As a result, these central parts become more metal-rich compared to the outer disk. While secular processes like radial migration tend to blur the gradient over time (Minchev et al. 2013), particularly in older stellar populations, our sample of early-type stars, representing a relatively young population, traces the current state of the disk.

The relationship between abundance and Galactocentric radius ($R_{\text{GC}}$) is shown in Fig. 14 for stars with $SNR_r > 50$ near the mid-plane ($|z| < 0.4$ kpc). BHB candidates ([Fe/H] < −1.4 dex) and chemically peculiar artifacts ([Fe/H] > 0.65 dex) are excluded. The remaining sample is grouped into radial bins, each containing at least 50 stars. Abundance values and uncertainties are computed as the weighted mean and standard deviation within each bin. The solid lines represent linear fits to our sample, with slopes indicated in the lower-left corner. Black dotted lines show Galactic gradients for each element derived from Classical Cepheids (Trentin et al. 2024).

The derived slopes and zero-points agree well with those obtained from Classical Cepheids, except for an offset in [O/H] (the second row), as discussed in Sec. 5.4. The close agreement between the gradients from our LAMOST early-type star sample and those from Cepheids reinforces the reliability of our methodology and supports the idea that recent chemical enrichment of the disk has been spatially and temporally coherent.

Furthermore, the abundance spreads remain homogeneous within $R_{\text{GC}} \leqslant 10$ kpc but increase significantly in the outer disk ($R_{\text{GC}} \geqslant 12$ kpc), particularly for [Fe/H] and [Si/H]. This trend aligns with the larger scatter reported in Cepheids at greater distances (Trentin et al. 2023), suggesting that the interstellar medium becomes less homogeneous farther from the Galactic centre.

Trentin et al. (2023) identified a possible break in the gradient at $R_{\text{GC}} = 9.25$ kpc, with slopes of $-0.063 \pm 0.007$ dex kpc$^{-1}$ and $-0.079 \pm 0.003$ dex kpc$^{-1}$ for the inner and outer regions, respectively. A similar trend is tentatively seen in [Fe/H], indicating a steeper gradient at larger distances. This consistency across different tracers and methods strengthens the interpretation that the observed gradient reflects the disk's star formation

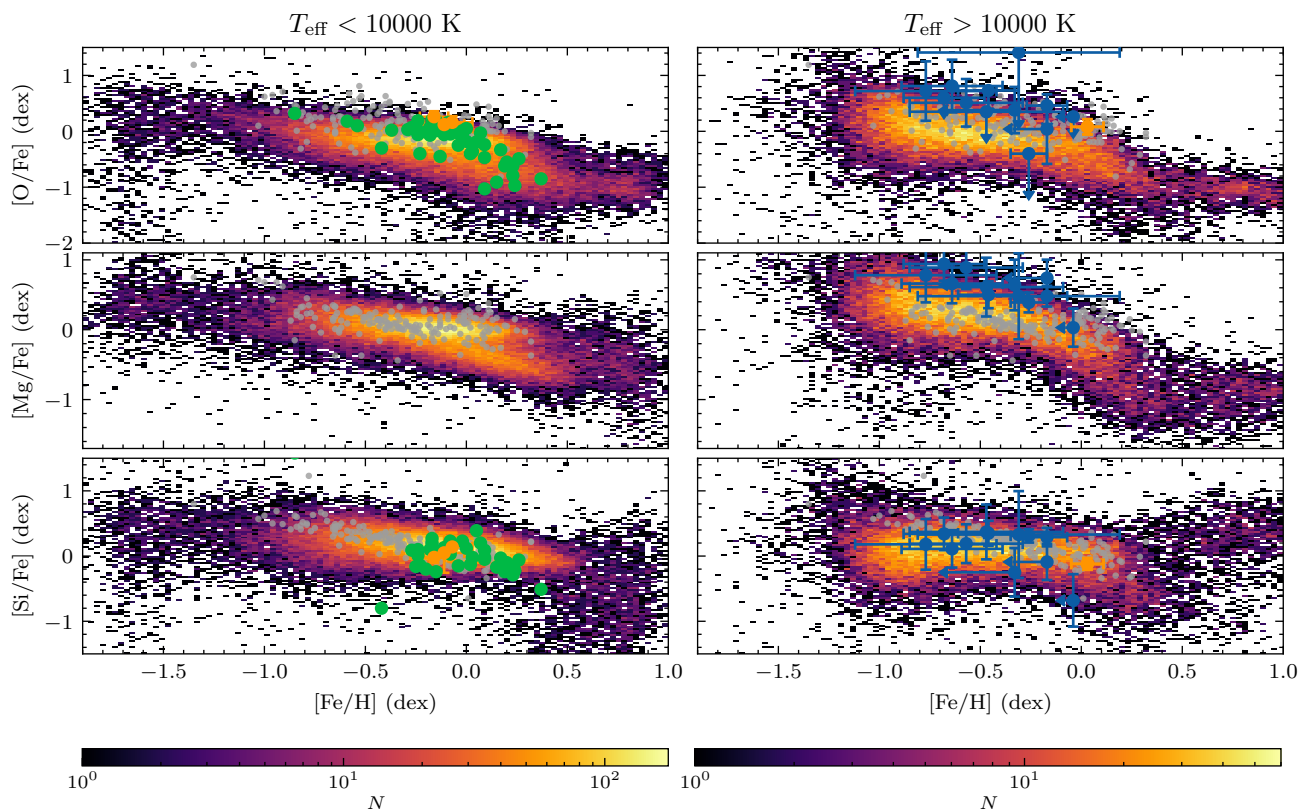

Fig. 13: Elemental abundance distribution of the sample in the [X/Fe] –[Fe/H] plane. Only stars with SNR$_r$ > 100 are shown. The sample is separated into two temperature bins: $T_{\mathrm{eff}} < 10000\,$K (left) and $T_{\mathrm{eff}} > 10000\,$K (right). Colored circles (with error bars where available) represent literature results from high-resolution spectroscopy used in Sec. 4.1, following the same colour scheme as Fig. 4. Grey dots represent the reference pattern of Classical Cepheids from Trentin et al. (2024).

history and dynamical evolution (but see also results from open clusters, e.g., Joshi et al. 2024).

# 6. Summary

In this work, we present stellar parameters ($T_{\mathrm{eff}}$, $\log g$, $v \sin i$) and abundances of four elements (O, Mg, Si, and Fe) for 315,822 stars with SNR ⩾ 10 in the $r$-band from LAMOST DR9 low-resolution spectra. We combine two spectroscopic analysis approaches, SLAM and the Payne, to determine parameters for hot stars ($T_{\mathrm{eff}} \sim 9500\,$K). The SLAM method, building on our prior work (Sun et al. 2021a,b; Sun & Chiappini 2024), demonstrates that stellar labels (particularly temperature and surface gravity) can be robustly derived from low- and medium-resolution spectra without excluding critical spectral features. Coupled with the Payne model, we extend this to abundance determinations for hot stars, offering new insights into massive star evolution and Galactic chemical enrichment.

Synthetic tests reveal improved precision in $T_{\mathrm{eff}}$ and $\log g$ recovery at higher SNR, though performance declines for hotter stars. Abundance measurements show temperature-dependent scatter, with [C/Fe], [S/Fe] and [Ca/Fe] excluded due to large uncertainties and weak SNR trends. Repeat observations yield typical uncertainties of 163 K in $T_{\mathrm{eff}}$, 0.12 dex in $\log g$, and 0.10 dex in [Fe/H], consistent with synthetic tests within a factor of two.

We verify labels against high-resolution benchmarks spanning $T_{\mathrm{eff}} \sim 8000$–19000 K and [Fe/H] $\sim -1.0$ to 0.5 dex, including three giants. Results show good agreement for $T_{\mathrm{eff}}$, $\log g$, and $v \sin i$. Iron abundances match literature values but diverge at $T_{\mathrm{eff}} > 15000\,$K, with metal-poor stars underestimated. [Mg/Fe] and [Si/Fe] show lowest scatter ($\sigma \sim 0.3$ dex), while [O/Fe] agrees moderately ($\sigma \sim 0.4$ dex). Comparisons with Xiang et al. (2022) and Sun et al. (2021a) reveal systematic offsets for 9000–11000 K stars, likely from differences in Balmer-line treatment.

Over 95% of the high-SNR sample satisfies photometric selection criteria from Zari et al. (2021), validating their utility for OBA-star identification.

We identify 3564 BHB candidates via metallicity ([Fe/H] ⩽ −1.4 dex) and Galactic latitude (|$b$| ⩾ 15°). 90% exhibit low $\log g$ (< 3.5 dex) and align with PARSEC horizontal-branch isochrones. A distinct subgroup (10%) with higher $\log g$ and moderate rotation ($v \sin i \sim 50\,\mathrm{km\,s^{-1}}$) likely represents blue stragglers. Discrepancies in $\log g$ with Xiang et al. (2022) highlight methodological differences.

Abundance trends for stars below 10000 K align with the high-$\alpha$ sequence from Hayden et al. (2015), while hotter stars show steeper slopes and elevated [O/Fe] and [Mg/Fe], suggesting temperature-dependent systematics. We identify a super metal-rich population ([Fe/H] > 0.5 dex) with narrow [O/Fe] scatter and a low-$\alpha$ sequence in [O/Fe] –[Fe/H] and [Mg/Fe] –[Fe/H] planes.

Our results confirm a negative abundance gradient consistent with previous studies of Classical Cepheids, with a slope of −0.070 ± 0.007 in [Fe/H] over Galactocentric distances between 6 and 15 kpc. This reinforces the reliability of our methodology and supports the notion that recent chemical enrichment in the

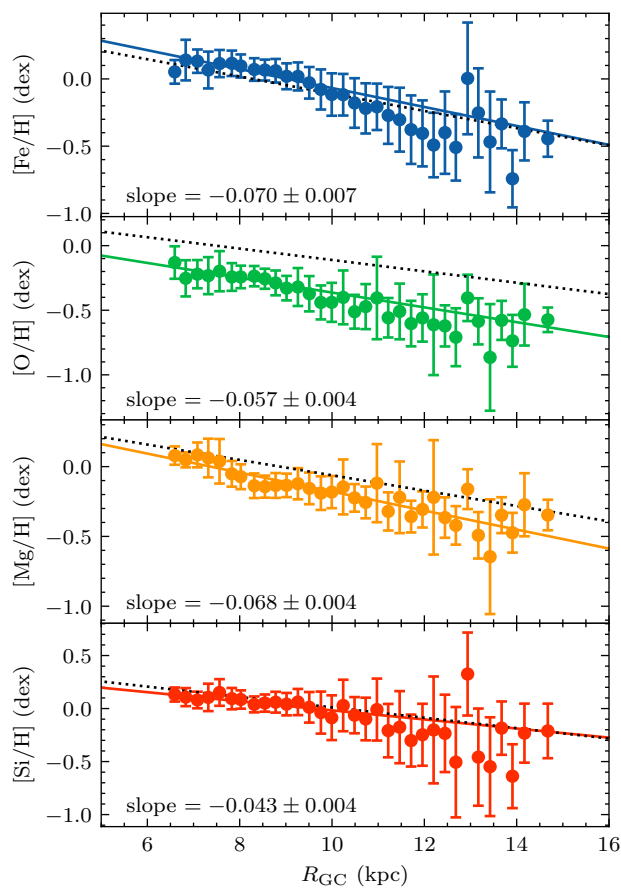

Fig. 14: Galactic radial gradients for [Fe/H], [O/H], [Mg/H], and [Si/H] (from top to bottom). Stars in the mid-plane ($|z| < 0.4$ kpc) with $SNR_r > 50$ are grouped into radial bins. Only bins containing at least 50 stars are shown, with values and uncertainties computed as the weighted mean and standard deviation of the abundances within each bin. The best-fit linear correlation is shown as a solid line, with its slope in the lower-left corner. The dotted line represents literature results for Galactic radial gradients derived from Classical Cepheids (Trentin et al. 2024).

Milky Way disk has been spatially and temporally coherent. The abundance spread remains uniform within $R_{GC} \leqslant 10$ kpc but increases significantly beyond 12 kpc, particularly in [Fe/H] and [Si/H], suggesting a less homogeneous interstellar medium at larger radii.

This catalogue establishes a foundation for studying Galactic hot stellar populations. Future work should incorporate NLTE modelling to address systematic offsets in [O/Fe] and [Mg/Fe] for rapidly rotating stars. Upcoming data releases SDSS-V (e.g., SDSS-V, Almeida et al. 2023) will enhance metal-poor samples where current deviations increase. Joint photometric-spectroscopic analyses could resolve log $g$ discrepancies for A-type stars via continuum constraints (e.g., Ju et al. 2025).

The synergy between SLAM and Payne frameworks demonstrates their adaptability for surveys like 4MOST (de Jong et al. 2019), and in particular, its 4MIDABLE-LR (Chiappini et al. 2019) consortium survey, enabling unified chemical inventories across stellar evolution stages. Such advances will critically advance our understanding of cosmic abundance variations and massive stars' role in Galactic chemical evolution.

## DATA AVAILABILITY STATEMENT

The data underlying this article are only available in electronic form at the CDS via anonymous ftp to cdsarc.cds.unistra.fr (130.79.128.5) or via http://cdsweb.u-strasbg.fr/cgi-bin/qcat?J/A+A/.

*Acknowledgements.* We are deeply grateful to C. Chiappini for the insightful discussions and comments. We thank the anonymous referee for their valuable comments. The Guoshoujing Telescope (the Large Sky Area Multi-Object Fibre Spectroscopic Telescope; LAMOST) is a National Major Scientific Project built by the Chinese Academy of Sciences. Funding for the project has been provided by the National Development and Reform Commission. LAMOST is operated and managed by the National Astronomical Observatories, Chinese Academy of Sciences. This work has made use of data from the European Space Agency (ESA) mission *Gaia* (https://www.cosmos.esa.int/gaia), processed by the *Gaia* Data Processing and Analysis Consortium (DPAC; https://www.cosmos.esa.int/web/gaia/dpac/consortium). Funding for the DPAC has been provided by national institutions, in particular the institutions participating in the *Gaia* Multilateral Agreement.
*Facility:* LAMOST
*Software:* astropy (Astropy Collaboration et al. 2013), IPython (Perez & Granger 2007), laspec (Zhang et al. 2021), matplotlib (Hunter 2007)

## Appendix A: Pixel-scale precision

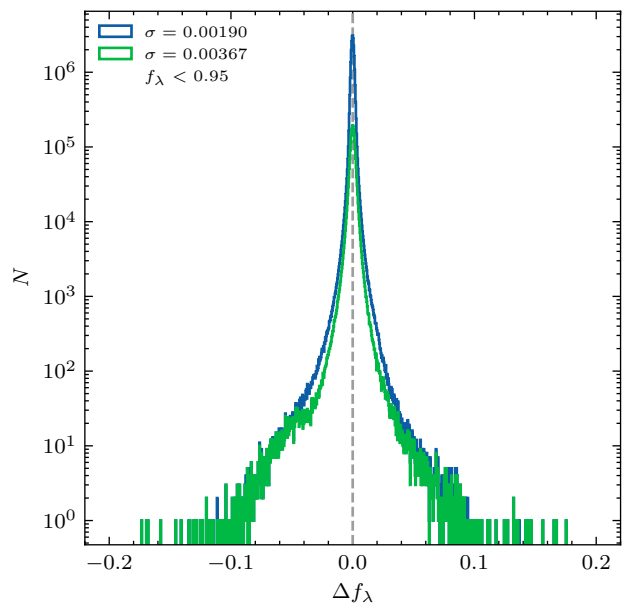

Fig. A.1: Distribution of flux residuals for 5000 test spectra at SNR = 1000. Standard deviations for all pixels (blue) and pixels with normalised flux $f_\lambda < 0.95$ (green) are indicated in the top-left corner. The latter criterion selects line-core regions of strong absorption features.

To evaluate the spectroscopic precision of our model (particularly the Payne implementation), we compare interpolated synthetic spectra with original test models at SNR = 1000. Fig. A.1 displays flux residual distributions for all pixels (blue) and line-core pixels with $f_\lambda < 0.95$ (green). The global pixel residual dispersion is 0.19% (equivalent to SNR ∼500 per pixel), significantly smaller than typical LAMOST-LRS spectral errors. This precision persists when examining H-line cores, where the dispersion increases to 0.37% (SNR ∼270) while remaining below observational uncertainties.

As Ting et al. (2017) demonstrated, gradient spectra $\nabla_l$ — quantifying spectral response to parameter variations — reveal spectroscopic information content. Fig. A.2 compares our Payne-derived gradients with Kurucz model calculations for a reference star with $T_{\rm eff} = 12500$ K. We compute gradients by differencing spectra with parameter offsets: $\Delta T_{\rm eff} = 200$ K, $\Delta \log g = 0.2$ dex, $\Delta v_{\rm mic} = 4\,{\rm km\,s^{-1}}$, $\Delta v \sin i = 50\,{\rm km\,s^{-1}}$, and $\Delta[{\rm Fe/H}] = \Delta[{\rm X/Fe}] = 0.2$ dex. The Payne gradients for $T_{\rm eff} = 12500$ K show excellent agreement with Kurucz models. Broad continuum features near strong absorption lines arise from normalisation effects, while sharp variations trace line profile changes.

We additionally compare a hotter reference star ($T_{\rm eff} = 17500$ K; orange in Fig. A.2) while maintaining identical other parameters, applying a vertical offset for clarity. The most notable trend is the shallower gradient amplitude in hotter stars compared to their cooler counterparts. This manifests in $\nabla T_{\rm eff}$ and $\nabla \log g$ through reduced variations near hydrogen line cores (Fig. A.2), a pattern similarly observed in $\nabla[{\rm Fe/H}]$ and $\nabla[{\rm X/Fe}]$. These diminished gradients imply temperature-dependent information loss: hotter stars' spectra contain weaker abundance sensitivity at fixed noise levels. While our model successfully reproduces the input gradient spectra, we emphasise that their absolute amplitudes ($< 2\%$) remain relatively flat compared to FGK stars. This information degradation intensifies with increasing temperature, underscoring the necessity for high-SNR ($\gtrsim 300$) observations even at LAMOST-LRS resolution to reliably constrain atmospheric parameters.

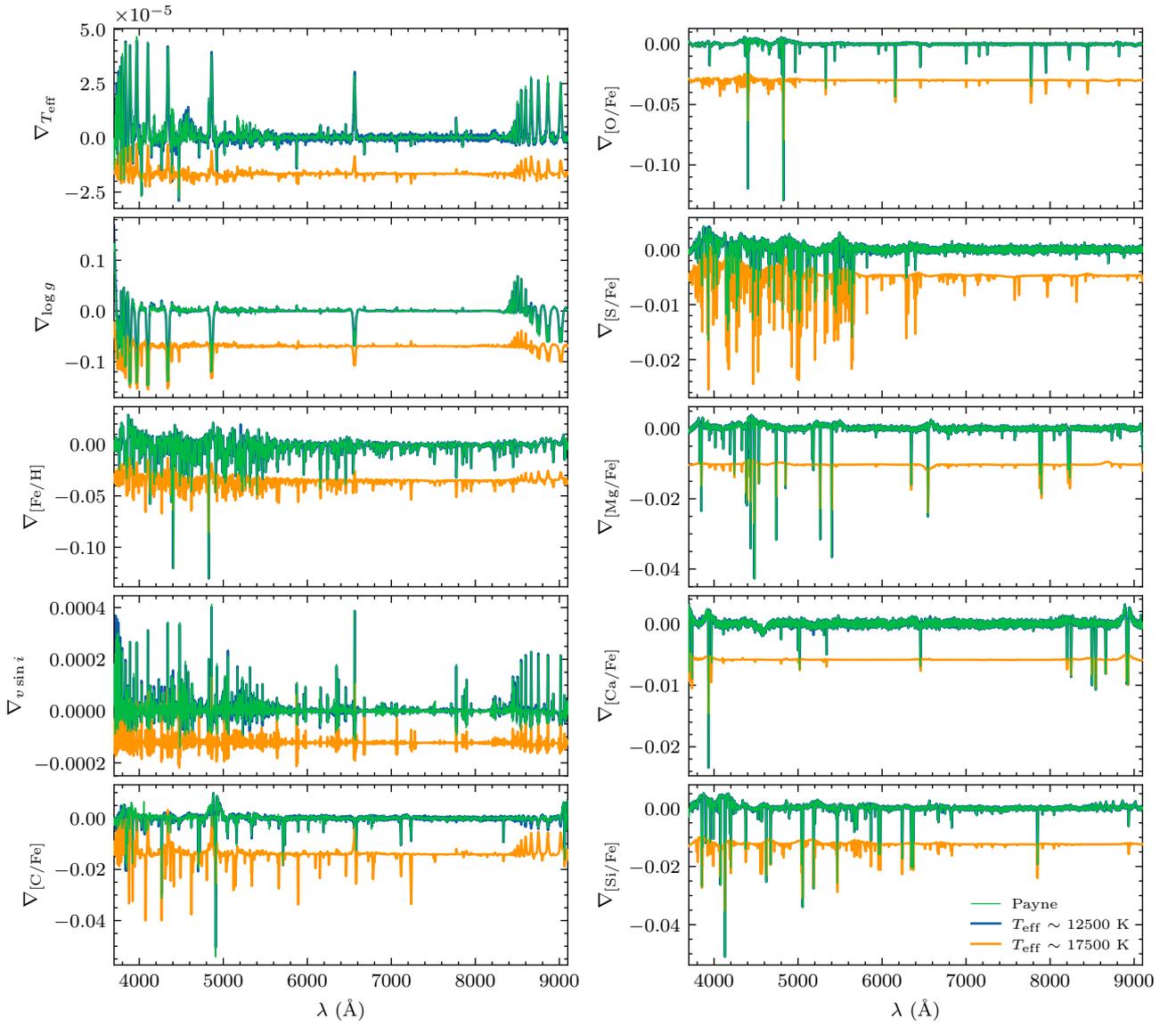

Fig. A.2: Gradient spectra $\nabla_l$ for reference stars with $T_{\rm eff} = 12500\,{\rm K}$ (blue) and $T_{\rm eff} = 17500\,{\rm K}$ (orange) from Kurucz models, compared to Payne model predictions (green) for $T_{\rm eff} = 12500\,{\rm K}$. A vertical offset separates the gradients for the hotter star. Both stars share $\log g = 3.5\,{\rm dex}$, $v_{\rm mic} = 5\,{\rm km\,s^{-1}}$, $v\sin i = 150\,{\rm km\,s^{-1}}$, [Fe/H] = 0.4 dex, and solar abundances for other elements.